\documentclass[11pt]{article}
\usepackage{amsmath}
\usepackage{amsfonts}

\usepackage{graphicx}
\usepackage{amsfonts,amssymb}

\newcommand{\al}{\ensuremath{\alpha}}
\newcommand{\ga}{\ensuremath{\gamma}}

\newcommand{\ka}{\ensuremath{\kappa}}
\newcommand{\la}{\ensuremath{\lambda}}

\newcommand{\om}{\ensuremath{\omega}}
\newcommand{\Om}{\ensuremath{\Omega}}

\newcommand{\Lag}{\ensuremath{{\cal L}}}
\newcommand{\Ricci}{\ensuremath{{\cal R}}}

\newcommand{\del}{\ensuremath{\partial}}

\newcommand{\half}{\frac{1}{2}}

\newcommand{\be}{\begin{equation}}
\newcommand{\ee}{\end{equation}}

\newcommand{\ba}{\begin{eqnarray}}
\newcommand{\ea}{\end{eqnarray}}

\newcommand{\ns}{\normalsize}

\newcommand{\eff}{\textrm{eff}}

\begin{document}
\begin{titlepage}

\title{
   \vskip 2cm
   {\Large\bf Cosmology with Twisted Tori}
\\[0.5cm]}
   \setcounter{footnote}{0}
\author{
{\ns\large 
  \setcounter{footnote}{3}
  Josef L. P. Karthauser$^1$\footnote{email: j.l.karthauser@sussex.ac.uk}~, 
  P. M. Saffin$^2$\footnote{email: paul.saffin@nottingham.ac.uk}~,
  Mark Hindmarsh$^1$\footnote{email: mark.hindmarsh@sussex.ac.uk}}
\\[0.5cm]
   $^1${\it\ns Department of Physics and Astronomy, University of
Sussex}\\
   {\ns Falmer, Brighton, BN1 9QJ, UK.} \\[0.2em] 
   $^2${\it\ns School of Physics and Astronomy, University of
Nottingham}\\
   {\ns University Park, Nottingham, NG7 2RD, UK.}
}

\maketitle

\begin{abstract}\noindent
We consider the cosmological role of the scalar fields generated
by the compactification of 11-dimensional Einstein gravity on a 7D
elliptic twisted torus, which has the attractive features of giving
rise to a positive semi-definite potential, and partially fixing
the moduli.  This compactification is therefore relevant for 
low energy M-theory, 11D
supergravity.  We find that slow-roll inflation
with the moduli is not possible, but that there is a novel scaling
solution in Friedmann cosmologies in which the massive moduli
oscillate but maintain a constant energy density relative to the
background barotropic fluid.
\end{abstract}

\thispagestyle{empty}
\end{titlepage}

\section{Introduction}



Current attempts to unify forces and interactions are mainly based on
superstring or M-theory with 10 or 11 space-time dimensions, with
the extra dimensions taking the form of a small compact manifold.
Finding such a manifold consistent with low-energy particle physics
and cosmology presents us with several hard problems.  A given
geometry has a number of size and shape parameters, or moduli, which
become scalar fields in the effective 4-dimensional theory, with a
potential generated in the first instance by the Ricci scalar of
the compact space. If the internal space is positively curved the
potential is negative, leading to anti de Sitter (AdS) 
solutions, which are disastrous in cosmological terms without extra ingredients such as
D-branes to cancel the negative cosmological constant \cite{Kachru2003a}.

Another problem with compactification is that some or all of the
moduli are not fixed by the effective 4-dimensional potential, at
least before quantum corrections are taken into account.  As these
moduli often affect parameters in the effective theory, we lose
predictability, and also gain unwanted massless scalars with
gravitational strength couplings.  The moduli which do appear in
the potential correspond to weakly coupled massive 4-dimensional
fields, which we must be careful not to excite, otherwise they will
dominate the energy density too early in a standard Friedmann
cosmology.

A promising route addressing some of these problems is to make use
of the flat-groups of Scherk and Schwarz \cite{Scherk1979}, which
have negative semi-definite curvature, and so lead to a theory with
a positive semi-definite potential and Minkowski minima. They also
fix some, but not all, of the moduli and so alleviate the moduli-fixing
problem generic to flat compactifications.

In this paper we study the cosmology of these flat-groups, or ``twisted
torus'' compactifications, with an eye to finding inflationary
behaviour and scaling regimes, in which the energy density in the
moduli fields remains a constant fraction of the background barotropic
fluid. The work forms a natural  continuation of the work presented
in \cite{Karthauser2006} where we studied the 7D cosets of 
compact Lie groups, classified in \cite{Castellani1984}.
In that case one finds that the curvature of these internal spaces is
unbounded and leads to a singular cosmology.
Applying the same technology to the particular class of twisted
tori introduced by Scherk and Schwarz we show firstly that slow-roll
inflation is not possible with the moduli fields (the no-go theorem
of \cite{Maldacena2001} does not apply here as the internal space
is allowed to be time-dependent).
We also find a novel scaling solution in which the massive moduli
are oscillating, but rather than dominating the fluid-filled
universe instead contributes a fixed fraction (of approximately 1/3
for radiation)
to the energy density.  The alleviation of the cosmological moduli
problem comes at a price: the volume of the internal space grows
approximately as $t^{0.5}$ during the radiation era.  As coupling
constants generically depend on the volume of the internal space,
a realistic cosmology exploiting this novel scaling regime appears
problematic.

The structure of the paper is as follows. Sections \ref{sec:dim:red:grav}
and \ref{sec:cosmScalar} are reviews of dimensional reduction of Einstein
gravity and scalar cosmology, with section \ref{sec:effTheory} giving 
the relevant formalism for multi-field cosmology and slow-roll behaviour.
Section \ref{sec:twisted} sees the introduction of twisted torus manifolds
which are then applied in the cosmological context in sections \ref{sec:canonical}
and \ref{sec:generic}. The evolutionary behaviour 
in terms of slow-roll and scaling is studied in sections
\ref{sec:slowRoll} and \ref {sec:scaling}, which we analyse in terms of effective
degrees of freedom in sections \ref{sec:effectiveScalarBehaviour} and \ref{sec:toyModel}.


\section{Dimensionally reduced gravity on a compact Lie manifold.}
\label{sec:dim:red:grav}

We start with pure gravity in $d+D$ dimensions which is described by the
Einstein-Hilbert action,
\ba
\label{eqn:biggravity}
\hat \Lag &=&\frac{1}{2\hat\kappa^2}\int d^d x \, d^D y \,
    \sqrt{-\hat g} \, \hat{\cal R},
\ea
where the hatted quantities are $d+D$ dimensional quantities.
We adopt a modified ansatz for the metric, given by the the Scherk-Schwarz
\cite{Scherk1979} form which guarantees a consistent truncation \cite{Pons2006},

\ba ds^2&=&e^{2\psi(x)}ds^2_{(d)}+g_{ab}(x) \nu^a(y) \otimes \nu^b(y)\\
\nu^a(y)&=&e^a(y)-{\cal A}^a(x)
\ea
where ${\cal A}^a$ are gauge fields and the $e^a$s are left-invariant
one-forms on the internal manifold, which we take to be a compact Lie group.
(Indices of the form $a,b,\dots$ span the $D$ extra dimensions.)
Upon substituting this into the Riemann scalar of the action (appendix 
\ref{AppReduceRicci}) we find an effective theory described by the scalars
$g_{ab}$, charged under the non-Abelian gauge fields ${\cal A}^a$.

Performing the $d^D y$ integration over the un-squashed volume of the internal
manifold allows us to make an identification of the form
\ba
\frac{1}{2\kappa^2} = \frac{{\cal V}_{\textrm{internal}}}{2\hat \kappa^2},
\ea
and gives a nice geometric interpretation of Newton's constant.  The constant 
$\kappa$ has the units of length and the $y$ co-ordinates are rescaled by
$\kappa$ such that they are dimensionless; this means that the $g_{ab}$
components have dimensions of $(length)^2$.

The determinant of the full metric decomposes into $\sqrt{-\hat g} =
e^{d\psi} \sqrt{- g_{(d)}} \sqrt{g_{ab}}$, and we can recover a pure $d$
dimensional
gravity term in the effective action by fixing the gauge of $\psi$ using
\ba
\label{psi:gauge:choice}
e^{(d-2)\psi}\sqrt{g_{ab}} = \kappa^D.
\ea

Specialising to $d=4$ and switching off the gauge fields (appendix \ref{app:ss:gauge:freedom}) we recover the effective
action
\ba
\label{eqn:effective:lagrangian}
\Lag_{\textrm{eff4}} 
 &=& \int d^4 x \sqrt{-g_4}
 \left[ \frac{1}{2\kappa^2}  \Ricci_4 - K^{abcd} \nabla_\mu g_{ab} \nabla^\mu g_{cd} -  V \right] \\
K^{abcd} &=& \frac{1}{2\kappa^2} ( \frac{1}{4} g^{ac} g^{bd} + \frac{1}{8} g^{ab} g^{cd}) \\
\label{effective:potential}
V &=& - \frac{1}{2\kappa^2} \frac{\kappa^D}{\sqrt{g_{ab}}} \Ricci_\textrm{internal},
\ea
where $K^{abcd}$ can be interpreted as a field space metric for the non-canonical
scalars $g_{ab}$, and $V$ is the potential in which they sit.  This is similar
to \cite{Hull2005} except that we don't make the restriction $\det(g_{ab}) = 1$
and thus avoid bringing in an additional scalar field.

For a general Lie group manifold the Ricci curvature can be shown to be \cite{Mueller-Hoissen1988}
\ba
\label{eqn:liegroup:ricci}
\Ricci_\textrm{internal} &=& \frac{1}{2} (f^c_{\;\;ab } \, 
        f^a_{\;\;dc }\, g^{db})
- \frac{1}{4}(f^c_{\;\;ab} \, f^f_{\;\;de } \, g^{ad} \, g^{be}\, g_{cf}),
\ea
where $f^a{}_{bc}$ are the structure constants \eqref{structureeqn} for the group,
and $g_{ab}$ is the choice of metric on the manifold.

\section{Cosmology with a scalar}
\label{sec:cosmScalar}

As we are interested in the cosmological dynamics of this theory with many
non-canonical scalars, we start by reminding ourselves of how the standard
single scalar picture works.

The simplest way of incorporating dark energy into a cosmological model is to
allow the negative pressure to be provided by a single scalar field,
\begin{align}
S &= \int \sqrt{-g}
    \left[
	\frac{1}{2\kappa^2}  \Ricci - \half \partial_\mu \phi \partial^\mu \phi - V(\phi)
    \right].
\end{align}
We take a Friedman-Robertson-Walker (FRW) universe
with homogeneity of the universe allowing the spacial gradients to be neglected,
and a barotropic fluid in the background with energy density $\rho_\gamma$ and
pressure $P_\gamma$ related by the equation of state
\ba
\label{eqn:eos:fluid}
P_\gamma = (\gamma-1)\rho_\gamma,
\ea
with $\gamma$
specifying the type of fluid ($\gamma = 4/3$ for radiation, $\gamma = 1$ for
pressureless matter and $\gamma = 0$ for vacuum energy). The equations of motion
for this system are
\begin{align}
\label{eqn:freidmann}
H^2 \equiv \frac{\dot a^2}{a^2} &=  \frac{\kappa^2}{3} \left[ \half \dot\phi^2
    + V + \rho_\gamma \right] \\
\label{eqn:acceleration}
\frac{\ddot a}{a} &=
    -\frac{\kappa^2}{3}
    \left[ \dot\phi^2 - V + \half(3\gamma -2)\rho_\gamma \right]\\
\label{eqn:kleingordon}
\ddot\phi+3H\dot\phi &= -\frac{dV}{d\phi}\\
\label{eqn:eom:rho}
\dot\rho_\gamma &= -3\gamma H \rho_\gamma.
\end{align}

Inflation is defined as a period of evolution where $\ddot a > 0$, by which
consideration we arrive at the so-called \emph{slow-roll} conditions. The
first of these, \cite{Steinhardt1984}
\ba
\epsilon \equiv \half\frac{1}{3\kappa^2}(V'/V)^2 &<<&1,
\ea
can be interpreted as saying that the slope is shallow enough for inflation
whilst the second,
\ba
\eta \equiv \frac{1}{3}\frac{V''}{\kappa^2V}<<1,
\ea
tells us that for inflation it should stay shallow for a while.

\section{Effective theory}
\label{sec:effTheory}

We can now compute the equations of motion of the theory of
Sec. \ref{sec:dim:red:grav}.  The degrees of freedom of the internal metric
become scalar fields under dimensional reduction, with an equation of motion
\ba
\label{eqn:gab:eom}
\ddot g_{ab} + 3 \frac{\dot a}{a} \dot g_{ab}
    &=& - \left(2g_{ac}g_{bd} -\frac{2}{D+2}g_{ab}g_{cd} \right) \kappa^2
    \del V/\del g_{cd} + g^{cd}\dot g_{ac} \dot g_{bd}
\ea
corresponding to \eqref{eqn:kleingordon}.
The scale factor equations of motion,
\ba
\label{eqn:friedman:effective}
H^2 \equiv \frac{\dot a^2}{a^2} &=& \frac{\kappa^2}{3} \left[ \half K
    + V + \rho_\gamma \right]
\\
\frac{\ddot a}{a} &=&
    -\frac{\kappa^2}{3}
    \left[ K - V + \half(3\gamma -2)\rho_\gamma \right]
\\
\dot\rho_\gamma &=& -3\gamma H \rho_\gamma,
\ea
take the same form as
(\ref{eqn:freidmann},\ref{eqn:acceleration},\ref{eqn:eom:rho}) where we define
\ba
\label{eqn:def:K}
K = K(g,\dot g) = 2 K^{abcd}\dot g_{ab} \dot g_{cd}
   = \frac{1}{4\kappa^2} (g^{ac} g^{bd}
    + \half g^{ab} g^{cd}) \dot g_{ab} \dot g_{cd},
\ea
which serves as the analogue of $\dot\phi^2$ in the single field case such
that the kinetic energy in the scalar sector is $\half K$. Expressed like
this it is apparent that although the theory is described by many non-canonical
scalars, the overall dynamics could be considered to be driven by a single
effective scalar with $\dot\phi_\textrm{eff} = \sqrt(K)$, at least
instantaneously.

We are also interested in the inflationary behaviour of the system, and
therefore it is useful to derive a form of the first slow-roll condition for this
kind of non-canonical system of scalars,
\ba
\label{eqn:noncanon:slowroll}
\epsilon \equiv &\frac{1}{6 V^2}\left[g_{ac}g_{bd} - \frac{1}{D+2}g_{ab}g_{cd}\right]
\frac{\del V}{\del g_{ab}}\frac{\del V}{\del g_{cd}} << 1.
\ea
With this we will be able to probe the field space numerically to
see whether there are any regions in moduli space in which inflationary
behaviour is possible.

\section{Twisted tori}
\label{sec:twisted}

In an earlier paper \cite{Karthauser2006} a model similar to
this one was studied where the internal manifold was chosen as
the 7D cosets of continous Lie groups $G/H$, classified in \cite{Castellani1984}.
It was  discovered there that these manifolds gave singular cosmologies due
to the ability of the curvature to change sign, and leading to potentials which
were unbounded from below.

We turn our attention now to the twisted tori manifolds of Scherk and Schwarz
\cite{Scherk1979}. These so-called ``flat-groups'' can be made compact by
forming a coset $G/\Gamma$ of a particular Lie group $G$ divided out by one of its
discrete subgroups $\Gamma$, and was shown there to allow a consistent
dimensional reduction of the type discussed here.
Most importantly for us however is that these cosets have negative
curvature leading to positive semi-definite potentials and therefore effective
theories with Minkowski minima.

These manifolds are related to the \emph{duality twists} of \cite{Hull2005,Catal-Ozer2006},
where they are identified as \emph{elliptic} twisted tori and related to the Lie
group $ISO(N)$. Two further groups which should fulfil the needs of our model
are also presented there, the \emph{hyperbolic} and \emph{parabolic} classes,
respectively related to the Lie group $SO(P,Q)$ and the Heisenberg group.  For
purposes of demonstration we consider here only the elliptic groups.

The elliptic groups are related to ISO(N), the isometries of a space of N
dimensions, in the following way. They are non-compact Lie groups with a Lie
algebra formed of N generators, $T_p$, corresponding to the translations and
$N(N-1)/2$ generators, $R_q$, corresponding to rotations. To form the elliptic
group the rotational
generators are coupled together to form a single generator $R= \sum_q m_q R_q$,
which when combined with the translational generators gives an $N+1$
dimensional subspace.  For such a group the structure constants are
given by,
\ba
\label{ttstructure}
f^i_{\;\;0j}&=&M^i_{\;\;j},
\ea
where $i,j=1,2,\dots,N-1$, and the $0$ direction corresponding to the rotational
generator. It is clear that this should be the case when one considers that
translations commute, and that the commutator of a translation and a rotation
will always be equivalent to a translation in a different direction. The resulting $M$ is
a skew symmetric, real $N$ x $N$ matrix populated by the $m_q$s, but we can
always choose a basis where $M$ takes the form
\ba
\label{eqn:twistedtori:M}
M&=&\left(
\begin{array}{ccccc}
0 & m_1 & 0    & 0 & 0\\
-m1 & 0 & 0    & 0 & 0\\
0   & 0 & 0    & m_2 & 0\\
0   & 0 & -m_2 & 0 & 0\\
0   &   &    0 & 0 & \ddots
\end{array}
\right).
\ea

This group is non-compact, and for the Kaluza-Klein ansatz to hold we
need a compact manifold, however this can be achieved by identifying
two separated points in each of the translation directions to form a torus, 
making a coset of the group with a discrete subgroup, ${\cal G}/\Gamma$ (see 
appendix \ref{app:elliptic:example}).  It is this resulting manifold which is the
elliptic twisted torus.

All that is left is to consider the form of the metric, $g_{ab}$, which is
permissable to put on the manifold. Unlike the coset manifolds of
\cite{Karthauser2006} there are no further symmetries with which to restrict the
metric, and so we are at liberty to consider all of the degrees of freedom in
our cosmological analyses. However there are still some gauge freedoms available
 which we can evoke to set $ g_{0i}=0 $ with
no loss of generality, and so we will work in this gauge from this point 
onwards.  We refer the reader to appendix \ref{app:ss:gauge:freedom} for
the details of this gauge fixing.

\section{Canonical scalars}
\label{sec:canonical}

For particle physics reasons we will assume that the internal manifold is seven
dimensional so that the entire space before compactification is of eleven
dimensions, descending from the pure gravity sector of 11D super-gravity, or
M-theory, for example. We are at liberty to place any metric we want on the
manifold, but for comparison with a system of canonical scalars we start our
investigation with the purely diagonal form
\ba
g_{ab} = \kappa^2 \textrm{diag}({A_0}^2, {A_1}^2, \dots, {A_6}^2),
\ea
substituting this into \eqref{eqn:liegroup:ricci} and finding that the
curvature is given by
\ba
\label{eq:curvature}
\Ricci = - \frac{1}{2\kappa^2 {A_0}^2} \left[
	\frac{ ({A_1}^2 - {A_2}^2)^2 }{{A_1}^2 {A_2}^2} {m_1}^2
    + \frac{ ({A_3}^2 - {A_4}^2)^2 }{{A_3}^2 {A_4}^2} {m_2}^2
    + \frac{ ({A_5}^2 - {A_6}^2)^2 }{{A_5}^2 {A_6}^2} {m_3}^2
    \right].
\ea
We can immediately see that as expected the curvature is
negative semi-definite, and this provides us via \eqref{effective:potential}
with a positive semi-definite potential, and a Minkowski minima for the theory. 
We
also observe that the metric structure follows that of the underlying geometry,
with the $A_0$ direction parametrising the rotational twist inherent in the
geometry, and the remaining directions pairing up into into three independent
planes in which the translation generators operate.

The metric is diagonal and so we are free to redefine these scalars such that we
normalise the kinetic terms and bring them into canonical form, which is a
useful exercise if only to compare the potential of the theory to other theories
that we may know. We therefore define a set of canonical scalars $\varphi_a$
related to the $A_a$s in the following way,
\begin{equation}
\begin{aligned}
A_0 &= e^{\kappa \,
	\left(
		\frac{1}{3} \sqrt{\frac{2}{7}}\, \varphi_{1}
		- \frac{\varphi_2}{\sqrt{2}}
		- \frac{\varphi_3}{\sqrt{6}}
		- \frac{\varphi_4}{2 \, \sqrt{3}}
		- \frac{\varphi_5}{2\, \sqrt{5}}
		- \frac{\varphi_6}{\sqrt{30}}
		- \frac{\varphi_7}{\sqrt{42}}
    \right) }
\qquad &
A_4 &= e^{\kappa \, \left(
	\frac{1}{3} \sqrt{\frac{2}{7}}\, \varphi_1
	+ \frac{2\, \varphi_5}{\sqrt{5}}
	- \frac{\varphi_6}{\sqrt{30}}
	- \frac{\varphi_7}{\sqrt{42}}
	\right) }\\
A_1 &= e^{\kappa \, \left(
    \frac{1}{3} \sqrt{\frac{2}{7}}\, \varphi_1
    + \frac{\varphi_2}{\sqrt{2}}
    - \frac{\varphi_3}{\sqrt{6}}
    - \frac{\varphi_4}{2\, \sqrt{3}}
    - \frac{\varphi_5}{2\, \sqrt{5}}
    - \frac{\varphi_6}{\sqrt{30}}
    - \frac{\varphi_7}{\sqrt{42}}
    \right) }
\qquad &
A_5 &= e^{\kappa \, \left(
	\frac{1}{3} \sqrt{\frac{2}{7}}\, \varphi_1
	+ \sqrt{\frac{5}{6}}\, \varphi_6
	- \frac{\varphi_7}{\sqrt{42}}
	\right) } \\
A_2 &= e^{\kappa \, \left(
	\frac{1}{3} \sqrt{\frac{2}{7}}\, \varphi_1
	+ \sqrt{\frac{2}{3}}\, \varphi_3
	- \frac{\varphi_4}{2\, \sqrt{3}}
    - \frac{\varphi_5}{2\, \sqrt{5}}
    - \frac{\varphi_6}{\sqrt{30}}
    - \frac{\varphi_7}{\sqrt{42}}
    \right) }
\qquad &
A_6 &= e^{\kappa \, \left(
	\frac{1}{3} \sqrt{\frac{2}{7}}\, \varphi_1
	+ \sqrt{\frac{6}{7}}\, \varphi_7
	\right) }. \\
A_3 &= e^{\kappa \, \left(
	\frac{1}{3} \sqrt{\frac{2}{7}}\, \varphi_1\
	+ \frac{{\sqrt{3}}\, \varphi_4}{2}
	- \frac{\varphi_5}{2\, \sqrt{5}}
	- \frac{\varphi_6}{\sqrt{30}}
	- \frac{\varphi_7}{\sqrt{42}}
	\right) }
\end{aligned}
\end{equation}
With these new scalars the effective potential and kinetic terms in
\eqref{eqn:effective:lagrangian} become,
\ba
\label{eqn:7dtt:canonpot}
V(\varphi_a) &=&
	\frac{1}{4\, {\kappa }^4} \,
    e^{-3\, {\sqrt{\frac{2}{7}}}\, \kappa \, \varphi_1 }
    \Bigg[
\\\nonumber&&
	  \qquad
	  \phantom{+}\;  e^{-2\, \sqrt{\frac{2}{3}} \, \kappa \, \varphi_{3}
	    + \frac{\kappa \, \varphi_4}{\sqrt{3}}
	    + \frac{\kappa \, \varphi_5}{\sqrt{5}}
	    + \sqrt{\frac{2}{15}} \, \kappa \, \varphi_6
	    + \sqrt{\frac{2}{21}} \, \kappa \, \varphi_7
	    }\,
	    \big( e^{\sqrt{2}\, \kappa \, \varphi_2}
	    	- e^{\sqrt{6}\, \kappa \, \varphi_3}
	    \big)^2 \, {m_1}^2
\\\nonumber&&
	\qquad
	+ \; e^{ \sqrt{2} \, \kappa \, \varphi_2
			+ \sqrt{\frac{2}{3}} \, \kappa \, \varphi_3
			- \frac{2\, \kappa \, \varphi_4}{\sqrt{3}}
	    	- \frac{4\, \kappa \, \varphi_5}{\sqrt{5}}
	    	+ \sqrt{\frac{2}{15}}\, \kappa \, \varphi_6
	    	+ \sqrt{\frac{2}{21}}\, \kappa \, \varphi_7
	    }\,
	    \big( e^{\sqrt{3}\, \kappa \, \varphi_4}
	    		- e^{\sqrt{5}\, \kappa \, \varphi_5} \big)^2\, {m_2}^2
\\\nonumber&&
	\qquad
    + \; e^{\sqrt{2}\, \kappa \, \varphi_2
    		+ \sqrt{\frac{2}{3}}\, \kappa \, \varphi_3
    		+ \frac{\kappa \, \varphi_4}{\sqrt{3}}
    		+ \frac{\kappa \, \varphi_5}{\sqrt{5}}
    		- 4\, \sqrt{\frac{2}{15}}\, \kappa \, \varphi_6
    		- 2\, \sqrt{\frac{6}{7}}\, \kappa \, \varphi_7}\,
    			\big( 
					e^{\sqrt{\frac{10}{3}}\, \kappa \, \varphi_6}
					- e^{\sqrt{\frac{14}{3}}\, \kappa \, \varphi_7}
				\big)^2\, {m_3}^2
		\Bigg] \\
\half K(\varphi_a) &=& \half \partial_\mu \varphi_a \partial^\mu \varphi_a,
\ea
and the theory is in canonical form.


\begin{figure}[!t]
\center
\includegraphics[width=8cm]{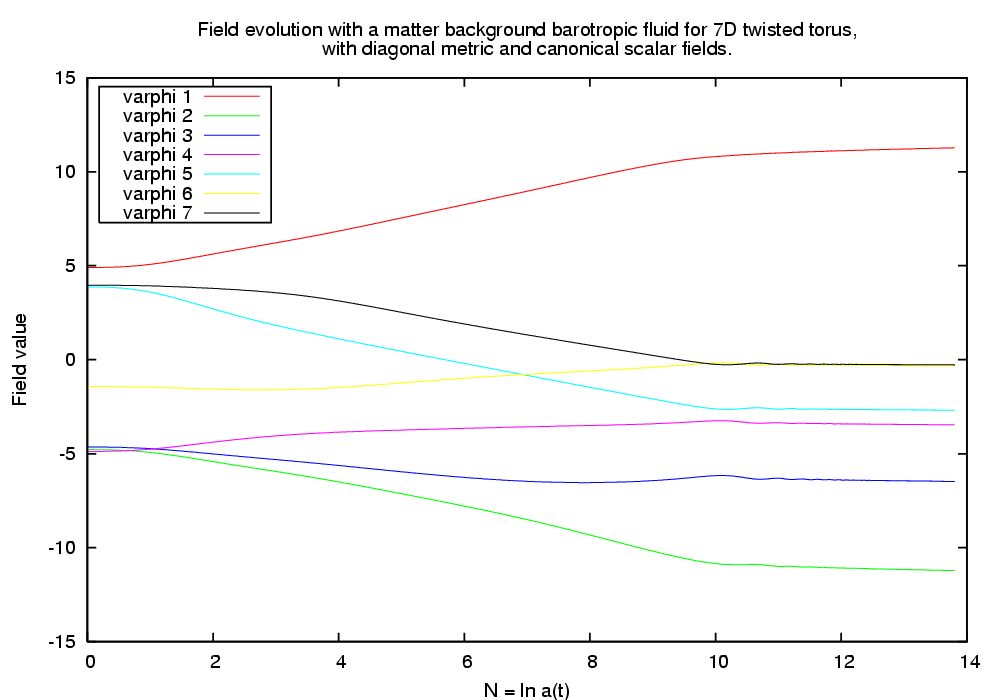}
\includegraphics[width=8cm]{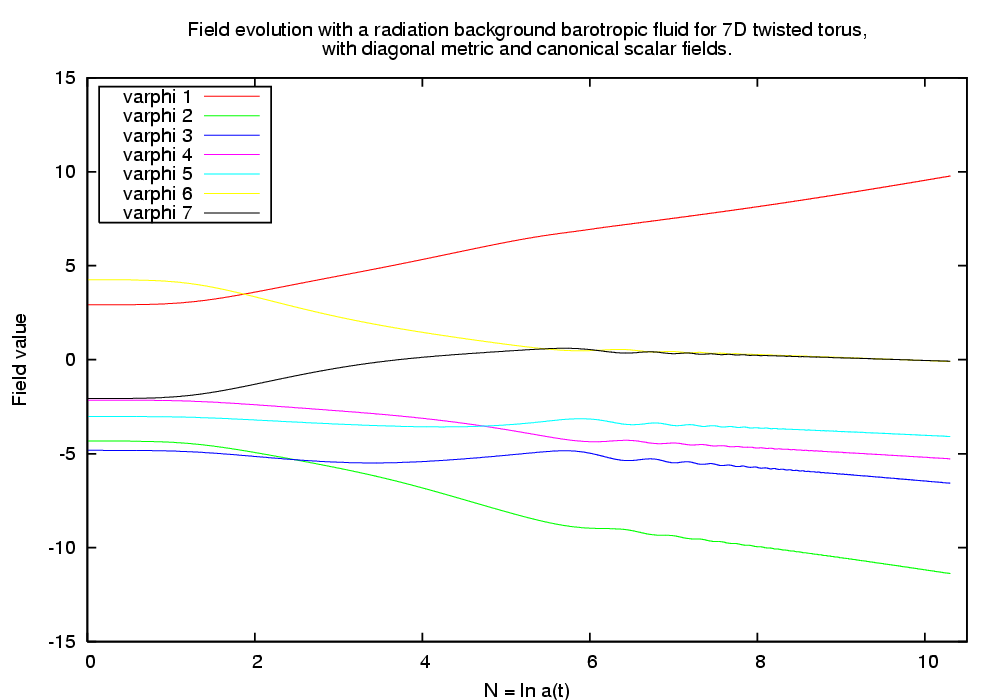}
\includegraphics[width=8cm]{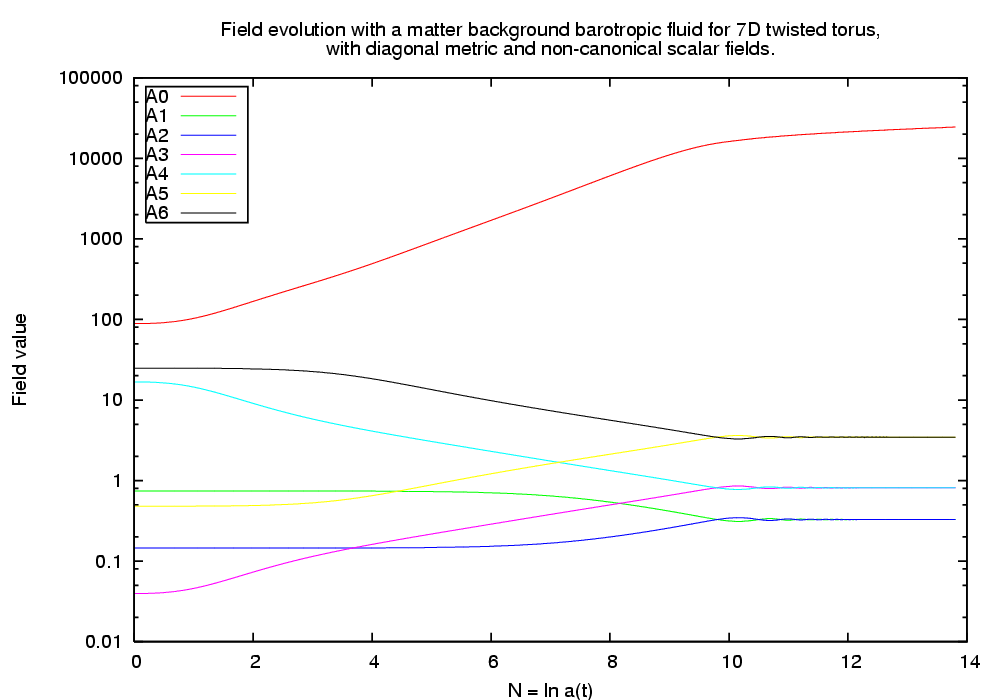}
\includegraphics[width=8cm]{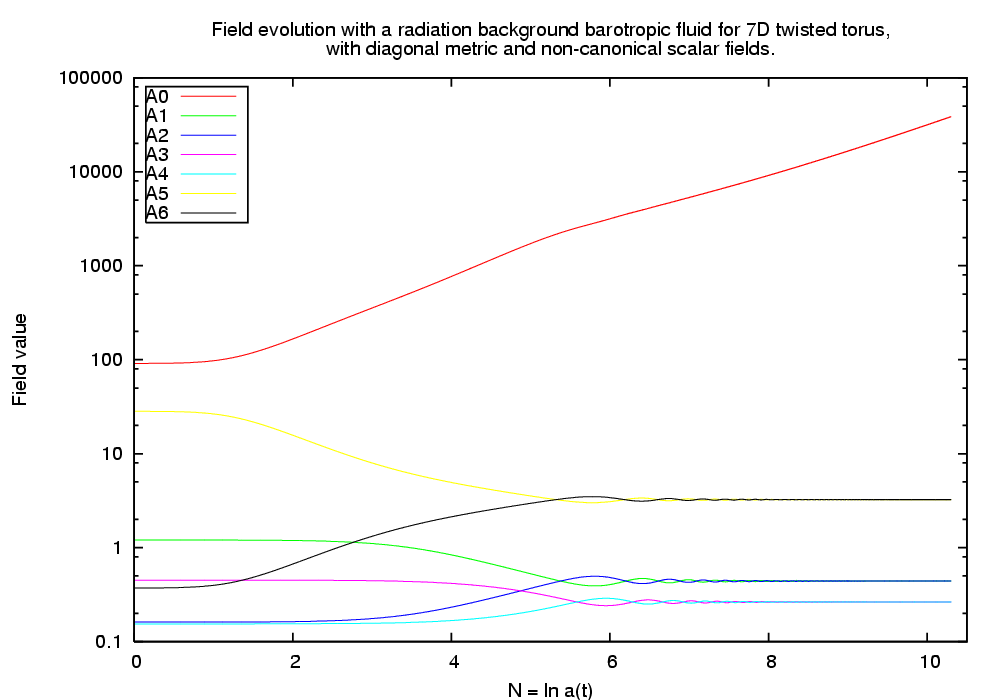}
\flushleft
\caption{Typical evolution of a system of canonical scalar fields evolving in
the effective potential of \eqref{eqn:7dtt:canonpot} along with a barotropic
fluid in the overall energy density. On the left is the evolution
with matter ($\gamma=1$) as the fluid component, and on the right with
radiation ($\gamma = 4/3$).  The upper plots are of the canonical scalars $\varphi_a$,
and the lower plots are of the non-canonical fields $A_a$, which are the
diagonal components of the internal metric. }
\label{fig:canon:fields}
\end{figure}

We ran numerical simulations with both matter and radiation present in the
barotropic fluid and a wide range of random initial field conditions, and
sampled from the results those which ran for the longest.
This was achieved using a fourth-order Runge-Kutta scheme, using the Friedmann
equation to monitor the accuracy of the simulation and allowing us to verify
that the dynamical variables had remained true to the constaint surface. In
Fig. \ref{fig:canon:fields} we present typical examples of the evolution of the
fields\footnote{We take $m_1 = m_2 = m_3 = 1$}.
One of the fields encodes for the volume of the
internal space which grows throughout the simulation, whilst the remaining
fields evolve in the fashion demonstrated until settling down into an
oscillation around the minimum of the potential at 
${A_1}^2 = {A_2}^2$, ${A_3}^2 = {A_4}^2$, ${A_5}^2 = {A_6}^2$.
    
The expansion of the volume demonstrates that the internal
manifold is in the process of decompactifying, which would suggest that as the
system evolves we would also need to consider non-zero Kalula-Klein modes in the
effective action.
However, in keeping with earlier studies \cite{Emparan2003} we shall not consider these modes,
and note that if 
decompactification happens slowly enough we can delay analysis of these
additional modes until much later in the history of the universe and avoid
spoiling any of results here.

\section{Non-canonical evolution}
\label{sec:generic}

As stated earlier this twisted torus places no restrictions on the form of the
internal metric, and
so we should say something about the evolution of the system with all
the degrees of freedom switched on.
The 7D internal metric has 28 degrees of freedom, of which 7 can be gauged away
(App. \ref{app:ss:gauge:freedom}). We choose to gauge away
6 of these allowing us to set
$g_{0i} = 0$, however that still leaves 22 scalar fields evolving
in the potential
\ba
\label{eqn:ncs:potential}
V = \frac{1}{2 \kappa^2} \frac{\kappa^D}{\sqrt{g_{ab}}}
	g^{00} (M^i{}_j M^j{}_i + M^i{}_k M^j{}_l g^{kl}\, g_{ij}).
\ea

\begin{figure}[!t]
\center
\includegraphics[width=8cm]{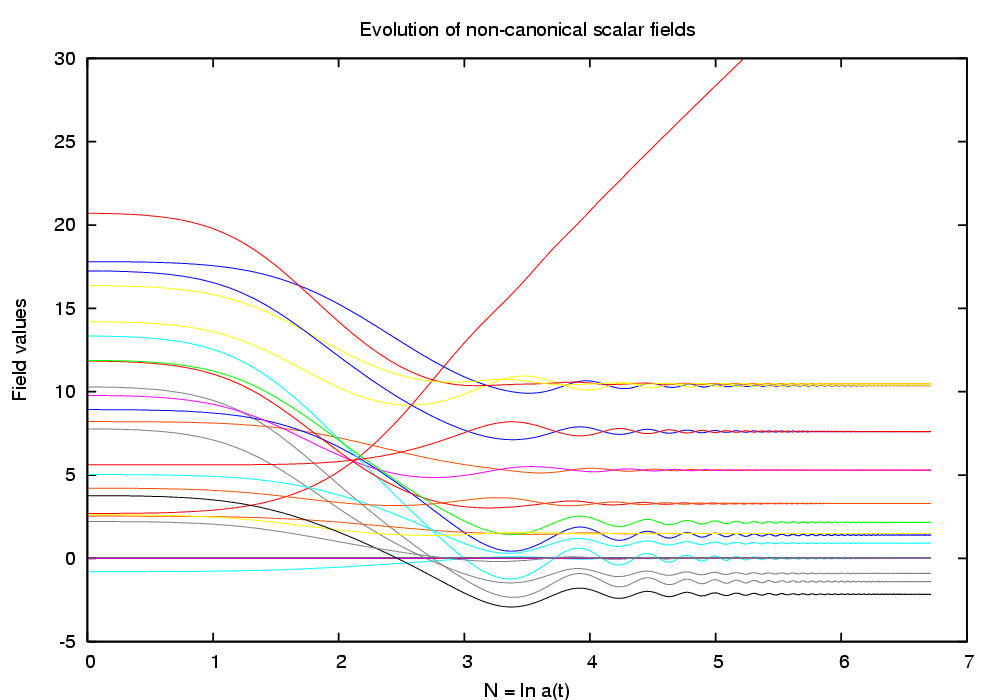}
\flushleft
\caption{Typical evolution of a system of non-canonical scalar fields, $g_{ab}$,
evolving according to equations
(\ref{eqn:effective:lagrangian}-\ref{effective:potential}).}
\label{fig:ncs:fields}
\end{figure}

A typical numerical realisation of this full non-canonical system is presented
in Fig. \ref{fig:ncs:fields}, in this case with matter in the background fluid.
As with the diagonal metric we see that all the scalars dynamically evolve
towards oscillating solutions where presumably the minimum of the potential
is reached, however this time the dependence of the potential upon the inverse
metric makes it difficult to determine where the minimum lies in field space.

\begin{figure}[!t]
\center
\includegraphics[width=8cm]{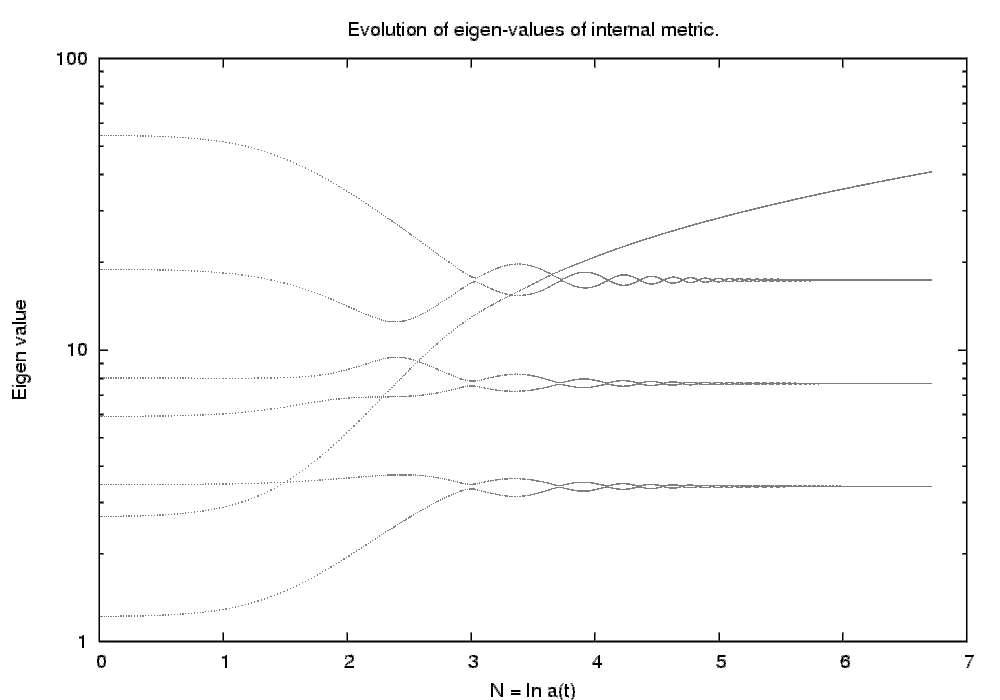}
\includegraphics[width=8cm]{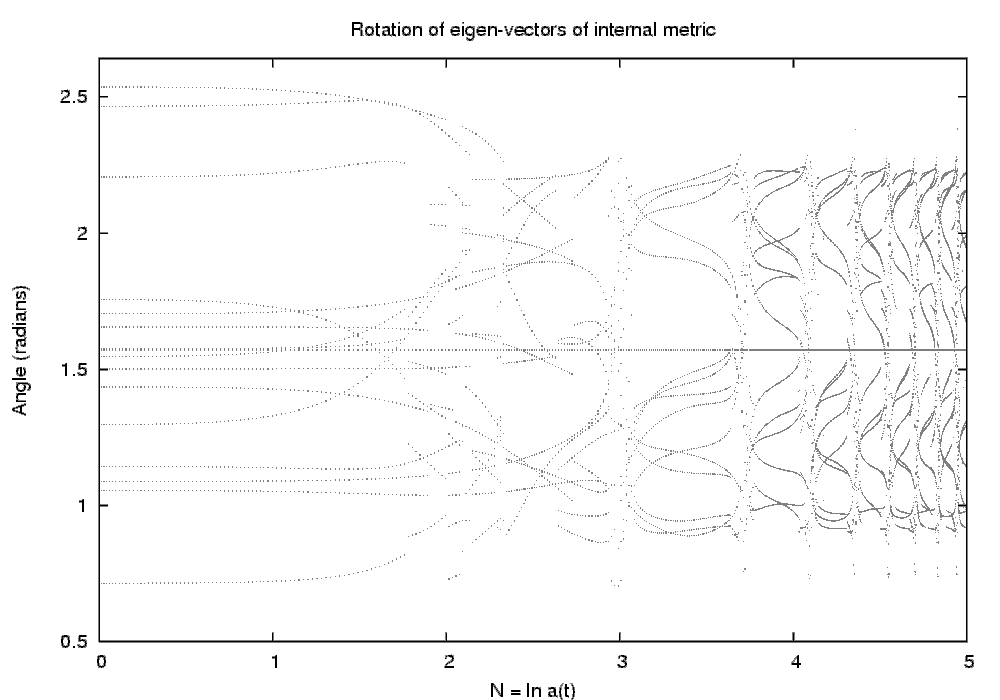}
\flushleft
\caption{Evolution of the eigensystem related to the scalar fields, $g_{ab}$.
On the left we have the eigenvalues, and on the right the angles of the
eigenvectors as measured relative to a canonical orthogonal frame ${\gamma_a}$.}
\label{fig:ncs:eigen}
\end{figure}

In order to make some sense of this we decomposed the step-wise values of the
metric into its associated eigensystem. To our surprise we discovered that the
evolution of the eigenvalues, as shown in the left-hand side of Fig. 
\ref{fig:ncs:eigen}, evolve in essentially the same fashion as the canonical
fields. Does this imply that only seven degrees of freedom are involved in the
dynamics?

As a step towards making some sense of this we remind ourselves how the fields
are related to the metric of the internal space. At a each point on a
manifold there is a frame of vectors, $\{e_a\}$, which form a basis for the
tangent space there.  The metric can be thought of as the ``dot product'' of these
vectors,
\ba
g_{ab} = e_a \cdot e_b = |e_a| |e_b| \cos \theta_{ab},
\ea
where $|e_a|$ is the length of the basis vector $e_a$ and $0<\theta_{ab}<\pi$
is the angle between the pair of basis vectors $e_a$ and $e_b$.  The diagonal
entries have $g_{aa} = |e_a|^2$ and so encode for the lengths of the frame
vectors, whereas the off-diagonal terms are also proportional to the cosine of
the angle between a pair of frame vectors.

Through the Kaluza-Klein ansatz we have integrated over the internal space
removing its co-ordinate dependence and so effectively there is only a single
point, a purely space-time one, at which this internal frame is located. The
scalar fields of the dimensionally reduced theory are the manifestion of the D
lengths and $D(D-1)/2$ angles between this set of D internal frame vectors.
That the evolution of the fields appears to be dependent upon just the
eigenvalues tells us only the lengths of the internal frame vectors appear
to play a role in the dynamics.  What then of the angles, what role do they play?

It is well known a real symmetric matrix is guaranteed to have
an orthogonal set of eigenvectors, that is every frame can be rewritten in terms
of a unique set of orthogonal vectors. We recover the angles associated with
the directions of these vectors via
\ba
\theta_{ab} \propto \cos^{-1} ( e_a \cdot \gamma_b )
\ea
where the $\{\gamma_a\}$ are a canonical set of orthogonal vectors satisfying
$\gamma_a \cdot \gamma_b = \delta_{ab}$, and we plot how these evolve in the
right-hand side of Fig. \ref{fig:ncs:eigen}.  It does appear that the frame is
rotating as the fields evolve, however we are unclear what role these rotations
play in the theory.

\section{Slow roll inflationary behaviour}
\label{sec:slowRoll}

We now turn our attention to the inflationary behaviour of the system.  We 
initially hoped that there might be a slow-roll regime, however the first
slow-roll condition \eqref{eqn:noncanon:slowroll} was never satisfied in any of
the numerical simulations. To understand this result we now turn our
attention to an analysis of the of the slow-roll condition for our
twisted torus. We start by
exploring the behaviour of the three-dimensional system with the diagonal metric,
$g_{ab} =
\kappa^2 diag(A^2, B^2, C^2)$. Expanding \eqref{eqn:noncanon:slowroll} we obtain
\ba
\epsilon  =
      1 + \frac{4}{3} \frac{\left( B^2 + C^2 \right)^2}{\left( B^2 -
      C^2 \right)^2},
\ea
which overtly fails the slow-roll condition and leads us to conclude that the
canonical three-dimensional system cannot ever slow-roll, which matches our
observations
under simulation. Further we find that adding the non-diagonal degree of
freedom,
\ba
g_{ab} = \left(
	\begin{smallmatrix} A^2 & 0 & 0 \\ 0 & B^2 & D^2 \\ 0 & D^2 & C^2
	\end{smallmatrix}
\right),
\ea
does not improve the situation.  Now slow roll is controlled by
\ba
\label{eqn:3d:nondiag:slowroll}
\epsilon  = 
  1 + \frac{4}{3}
  	\frac{\left(B^2 + C^2 \right)^2}{\left( \left( B^2 - C^2 \right)^2 + 4\, D^4 \right) },
\ea
for which we also have $\epsilon \geq 1$ and so once again there is no slow-roll
regime here either, and the additional degree of freedom does not change the
picture significantly.

The same pattern appears with the five-dimensional system;
with $g_{ab} = \kappa^2 diag(A^2, B^2, C^2, D^2, E^2)$ we find
\ba
\epsilon = 1 + \frac{4}{3} \frac{\left( {\left( B^4 - C^4
\right) }^2\,D^4 E^4\,{m_1}^4 + 
       B^4\,C^4\,{\left( D^4 - E^4 \right) }^2\,{m_2}^4 \right) }{
     {\left( {\left( B^2 - C^2 \right) }^2\,D^2\, E^2\,{m_1}^2 + 
         B^2\,C^2\,{\left( D^2 - E^2 \right) }^2\,{m_2}^2 \right) }^2},
\ea
where the second term is once again overtly positive definite.
Finally the diagonal seven-dimensional system of Sec. \ref{sec:canonical} with metric
$g_{ab} = \kappa^2 diag(A^2, B^2, C^2, D^2, E^2, F^2, G^2)$ reveals this pattern as
being generic,
\begin{align}
\epsilon  =& 
  1 + \frac{4}{3} \\
  & \frac{\left( F^4\, 
            G^4\, \left(
            	\left( B^4 - C^4 \right)^2\, D^4\, E^4\, {m_1}^4
            	+ B^4\, C^4\, \left( D^4 - E^4 \right)^2\, {m_2}^4 \right)
            	+ B^4\, C^4\, D^4\, E^4\, \left( F^4 - G^4 \right)^2\, {m_3}^4 \right) }
		{\left( F^2\, 
                G^2\, \left( \left( B^2 - C^2 \right)^2\, D^2\, E^2\, {m_1}^2
                + B^2\, C^2\, \left( D^2 - E^2 \right)^2\, {m_2}^2 \right)
                + B^2\, C^2\, D^2\, E^2\, \left( F^2 - G^2 \right)^2\, {m_3}^2 \right)^2}.\nonumber
\end{align}

We predict that a full non-canonical system
with generic non-diagonal metric will also fail to slow-roll, pointing
to the results of Sec. \ref{sec:generic} and
\eqref{eqn:3d:nondiag:slowroll} as evidence, and conclude that with elliptic
twisted tori in the internal manifold  there can be no slow-roll inflation.

\section{Scaling solutions}
\label{sec:scaling}

One of our motivations for studying the model of this paper was to determine whether
this choice of internal manifold could permit regimes of scaling behaviour in
the effective theory, and so it is to this question that we now turn.

To study such behaviour it is usual to reformulate the equations of motion in
terms of an automonous system \cite{Copeland1998,Collinucci2005,Hartong2006}.
The Friedmann equation \eqref{eqn:friedman:effective} describes the energy of
the system and can be recast in terms of the variables,
\ba
\label{eqn:aut:vars}
x \equiv \frac{\kappa \sqrt{K}}{\sqrt{6}H} \qquad \qquad
	y \equiv \frac{\kappa \sqrt{V}}{\sqrt{3}H} \qquad \qquad
	z \equiv \frac{\kappa \sqrt{\rho_\gamma}}{\sqrt{3}H}
\ea
which become constant in a scaling regime, and in terms of these the expression
of energy conservation can be rewritten more simply as
\ba
    x^2 + y^2 + z^2 = 1.
\ea

In a cosmological scaling solution the various components of the energy density
evolve in constant ratio to each other, causing the effective equation of state
parameter in the scalar sector,
\ba
\label{eqn:eosp:scalars}
\gamma_\varphi \equiv \frac{\rho_\varphi + P_\varphi}{\rho_\varphi}
    = \frac{2K}{K+V} = \frac{2x^2}{\Omega_\varphi}
\ea
to track that of the background fluid \eqref{eqn:eos:fluid}, where
$\rho_\varphi$ and $P_\varphi$ are respectively the energy density and pressure
in the scalar sector, and $\Omega_\varphi \equiv \kappa^2 \rho_\varphi / 3H^2
= x^2 + y^2$ is the total energy
density of the scalar sector. For these kinds of systems the evolution of the
scale factor can be shown to evolve according to $a(t) \propto t^p$, where
$p = 2/3\gamma_\varphi$.

\begin{figure}
\center
\includegraphics[width=8cm]{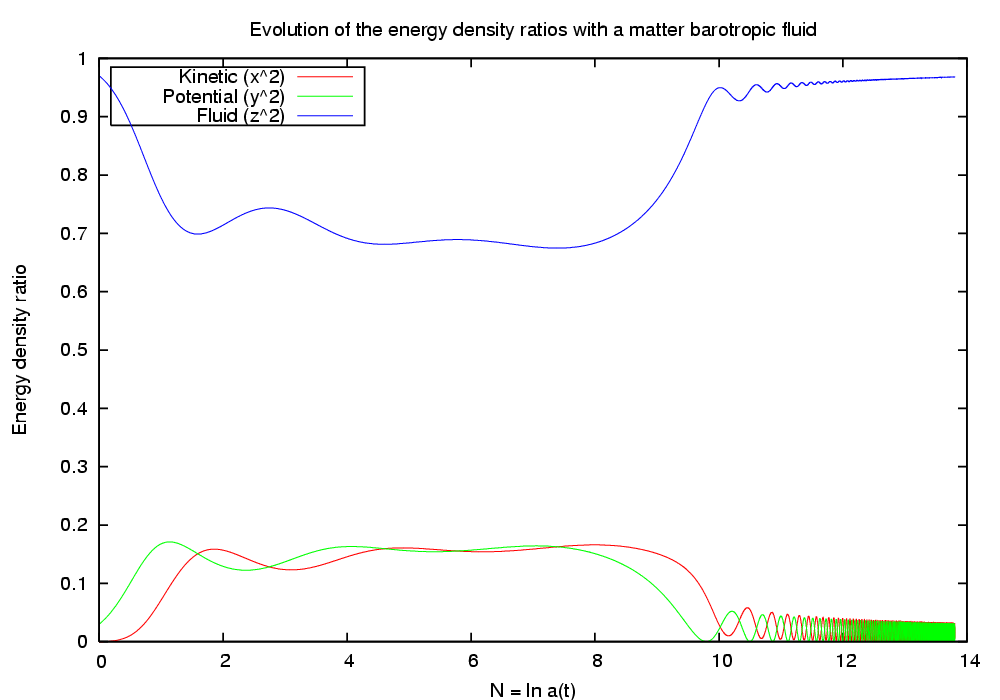}
\includegraphics[width=8cm]{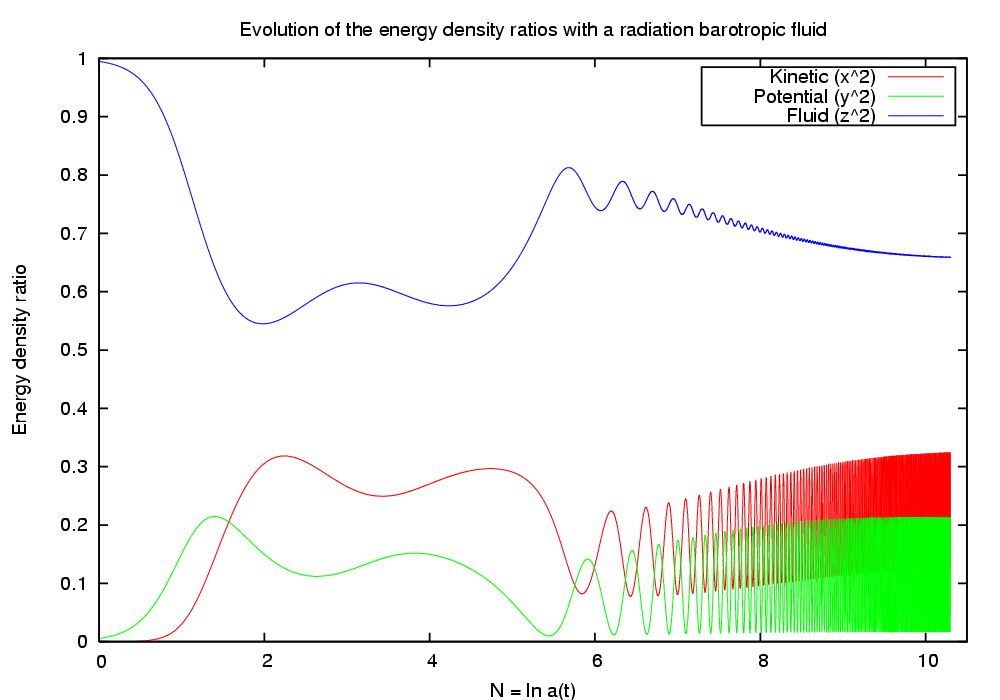}
\includegraphics[width=8cm]{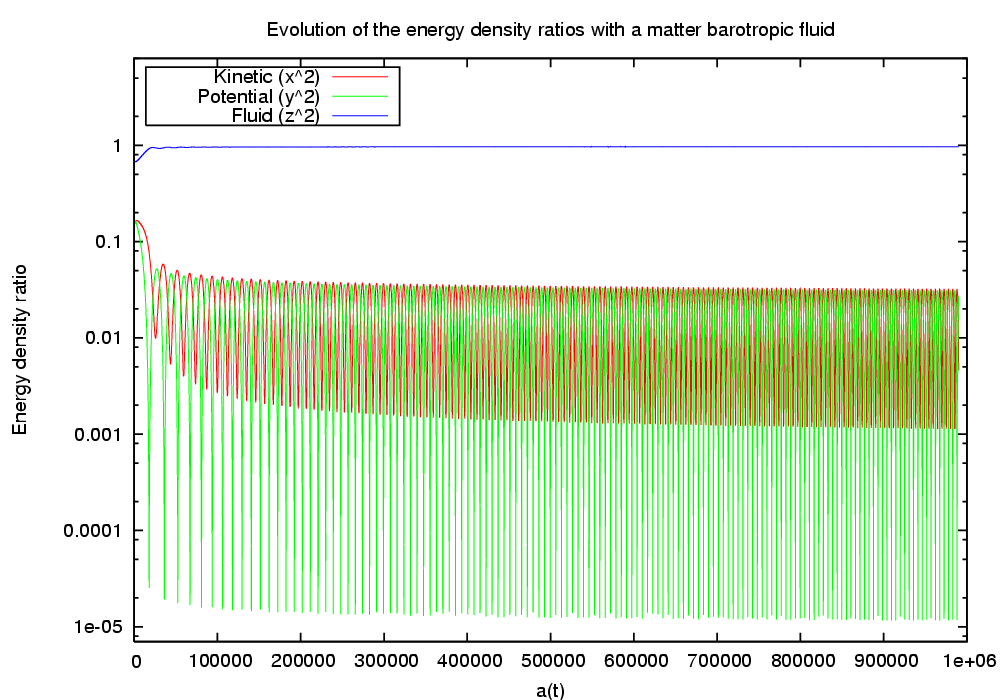}
\includegraphics[width=8cm]{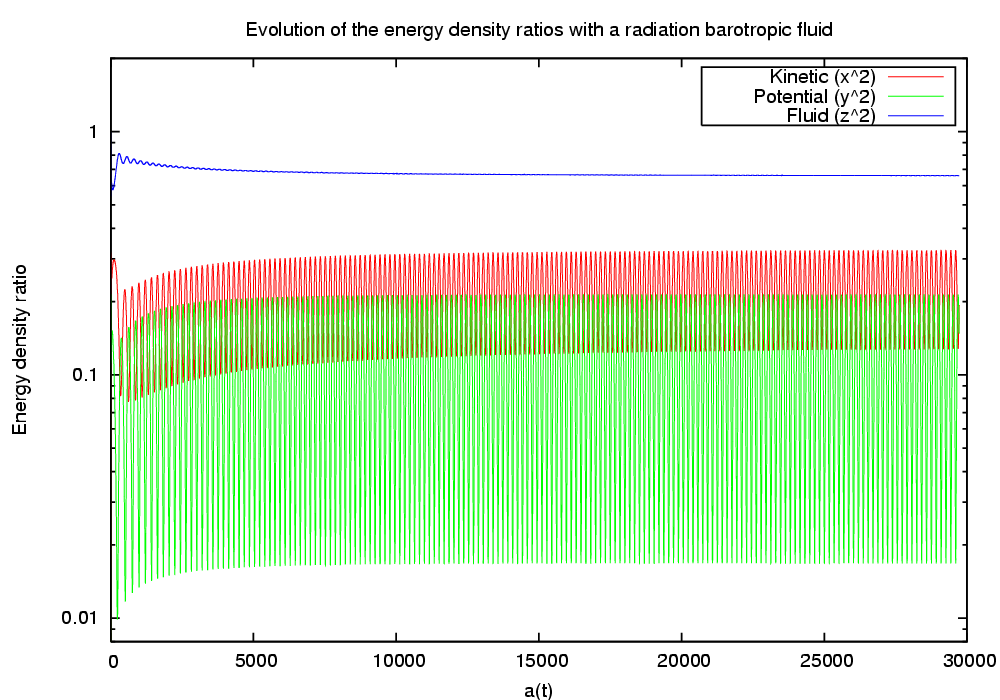}
\flushleft
\caption{Evolution of the energy densities of the components of
the fluid, for the simulations presented in Fig. \ref{fig:canon:fields}.
On the left is the evolution with matter ($\gamma=1$) as the fluid component, and
on the right that with radiation ($\gamma = 4/3$). Comparison between the top
and bottom graphs show that initially there is a period of scaling before an
oscillating phase takes over.}
\label{fig:canon:scaling:components}
\end{figure}

To see whether the system exhibits scaling we need to study the
evolution of the autonomous system variables \eqref{eqn:aut:vars}.
In Fig. \ref{fig:canon:scaling:components}
we have a plot of these scaling quantities for the evolutions of Fig. \ref{fig:canon:fields},
and we do indeed see that the system settles into what
looks like a scaling regime, at least for a short period, before transitioning
into a long phase where the energy density components appear to be oscillating
around constant values.  This oscillating behaviour appears to continue
generically with the amplitude reducing very slowly, and it does therefore
appear to be in an \emph{effective scaling} regime, oscillating around a true
scaling solution. We also note the plots show that each
contribution to the energy density is nowhere vanishing.

\begin{figure}
\center
\includegraphics[width=8cm]{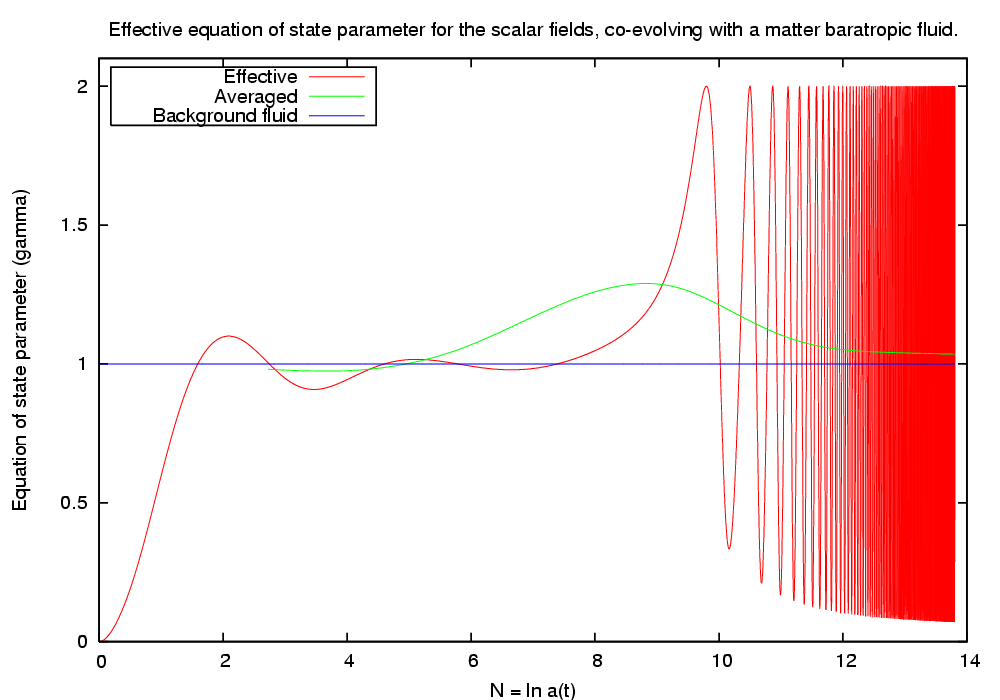}
\includegraphics[width=8cm]{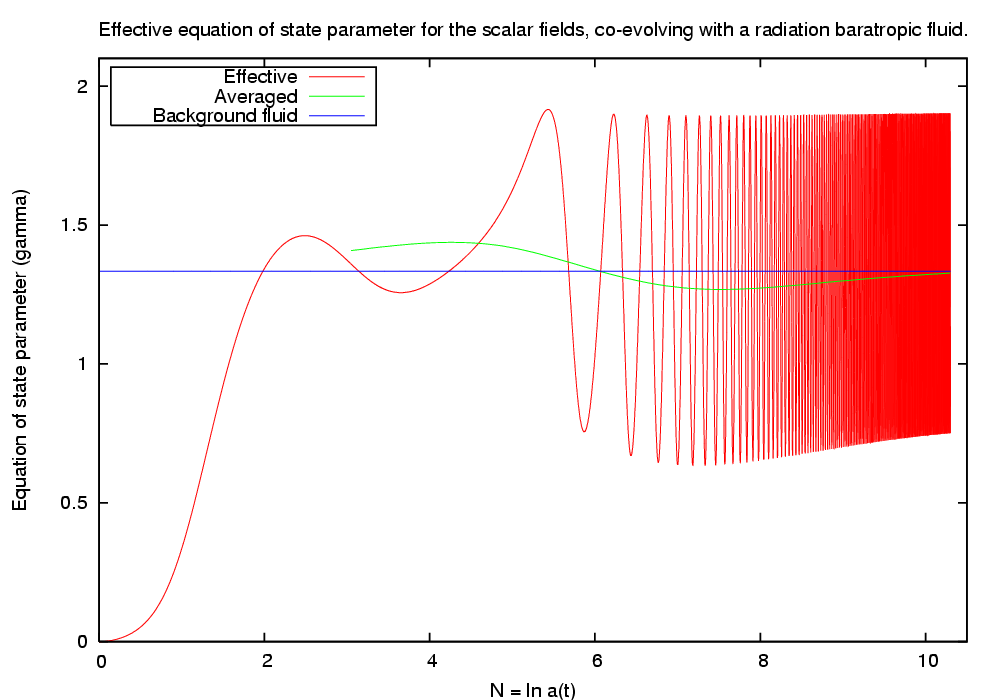}
\flushleft
\caption{The evolution of the equation of state parameter for the scalar fields,
given by $\gamma_\varphi = \frac{2K}{K+V}$, is shown along with that of the
background baratropic fluid.  The mid-point of $\gamma_\varphi$ during the
oscillating phase is also shown, which clearly shows that the oscillations
asymptote to same equation of system of the fluid.  This suggests that the
system is scaling.}
\label{fig:canon:eos}
\end{figure}

Further evidence that the system is in scaling can be seen from a plot of the 
equation of state parameter for the scalar field \eqref{eqn:eosp:scalars}. It 
is well known from the autonomous system analysis \cite{Copeland1998} that 
in the fluid dominated case there is a tracker solution in which the equation 
of state parameter of the scalar sector mimics that of the background 
barotropic fluid. In Fig. \ref{fig:canon:eos} we see $\gamma_\varphi$ does 
indeed appear to oscillate around $\gamma$, with the centre of oscillation 
appearing to approach the value of the background fluid asymptotically.  We 
take this as further evidence that this system of scalars is effectively 
scaling.

This behaviour was found to be characteristic of this system of scalars; in all
numerical simulations there was period where the system appeared to converge
on a scaling solution which, although pronounced in the data presented here, was
present in all the evolution runs we examined. This was always followed by a
phase
change into oscillating behaviour around another scaling solution. It appears
that this generic behaviour with non-vanishing potential and kinetic energy
could carry on for some time, although in our
runs we had to terminate the simulations early as the oscillations lead to exponential
memory requirements and limited the ability to run for much longer than the
examples shown here.

\section{Effective scalar behaviour}
\label{sec:effectiveScalarBehaviour}

Many workers have studied scaling behaviour in cosmologies with scalar fields
and exponential potentials \cite{Copeland1998,Collinucci2005,Hartong2006}.
Our potential \eqref{eqn:7dtt:canonpot} is also of exponential form but with too
many
terms to be understood in terms of these analyses, however it may be possible to
find an effective potential which broadly behaves in the same manner. This is
motivated by the observation that in \eqref{eqn:friedman:effective} the quantity
$\sqrt K$ could be considered to be analogous to 
the time derivative of an effective scalar field
$\dot \phi_\eff$. Can the dynamics alternatively be described in terms of
the evolution of this effective scalar, and if so what form might its
effective potential take?

\begin{figure}[!t]
\center
\includegraphics[width=8cm]{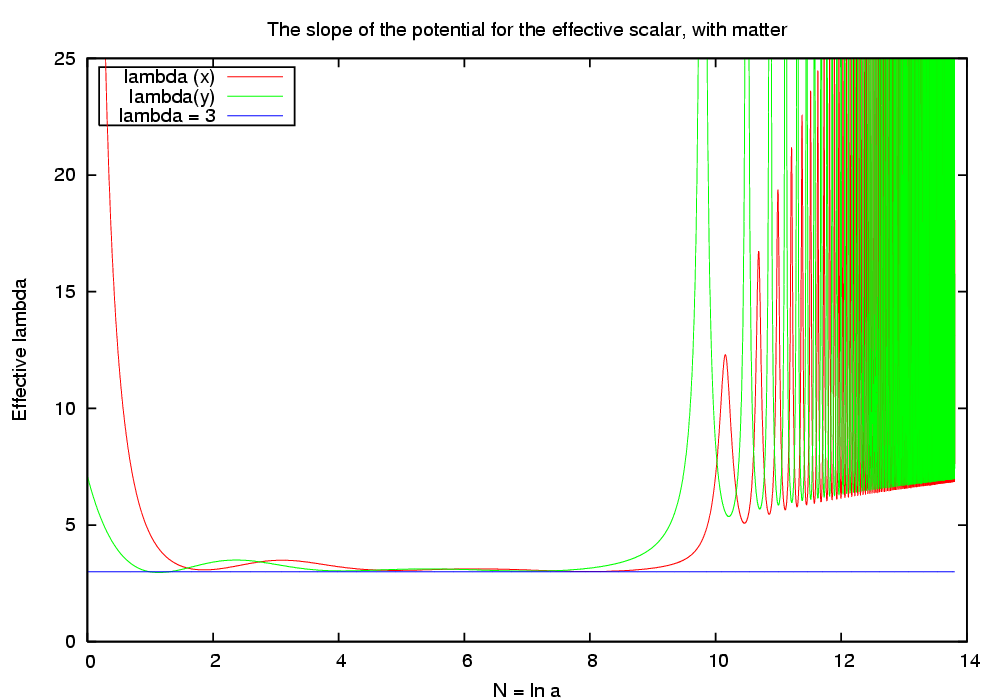}
\includegraphics[width=8cm]{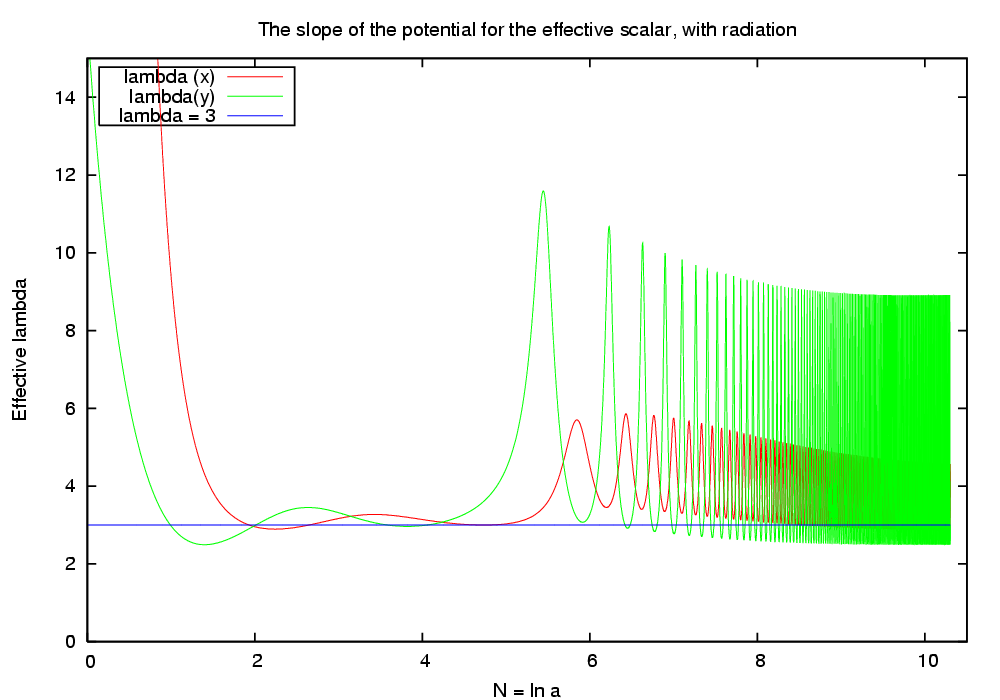}
\flushleft
\caption{Values of the slope $\lambda_\eff$ of the effective potential
\eqref{eqn:simple:exp:pot}, calculated in terms of the the autonomous system
variables \eqref{eqn:aut:vars} and the effective equation of state parameter for
the scalar fields \eqref{eqn:eosp:scalars}. }
\label{fig:canon:scaling:lamda}
\end{figure}

The simplest exponential potential of a single field that we can write down is
\ba
\label{eqn:simple:exp:pot}
V_\eff(\phi_\eff) = V_0 e^{-\lambda_\eff \phi_\eff},
\ea
of which the asymptotic scaling behaviour is well known. We see from
Fig. \ref{fig:canon:scaling:components} that the energy density is entirely dominated
by the fluid component, for which analytically an attractor solution for the
effective scalar exists with $\gamma_\varphi = \gamma$.
Assuming the dynamics of the full system is caught in this attractor we are
lead directly to a value for $\lambda_\eff$; in terms of the autonomous system
variables $x$ and $y$ we find
\ba
\label{eqn:eff:lambda}
\lambda_\eff
	= \sqrt{\frac{3}{2}} \frac{\gamma_\varphi}{x}
	= \sqrt{\frac{3}{2}}\frac{\sqrt{(2-\gamma_\varphi)\gamma_\varphi}}{y},
\ea
which we show calculated both ways in Fig. \ref{fig:canon:scaling:lamda}.

The single field approximation appears to be a fairly good one,
at least initially,
with the system evolving around a scaling solution with $\lambda_\eff \approx 3$
before the transition into the oscillating phase.  After this transition however
it is unclear what this approach can tell us; with radiation it appears that there
could be some effective lambda value, probably different from the initial one,
that the system is evolving around asymptotically, however this is less clear
with matter.

\begin{figure}
\center
\includegraphics[width=8cm]{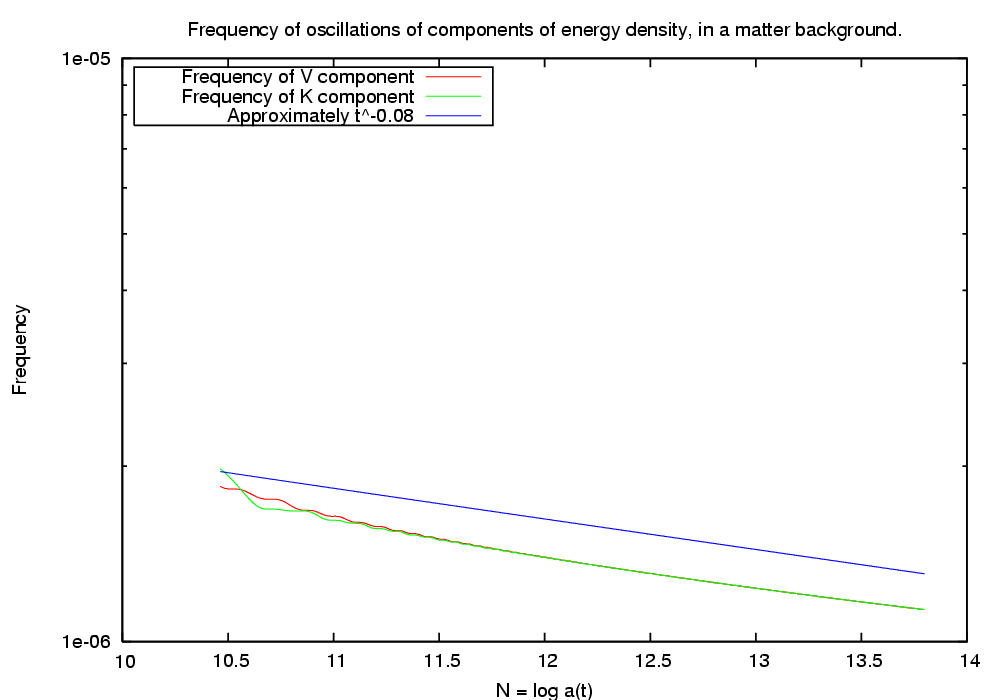}
\includegraphics[width=8cm]{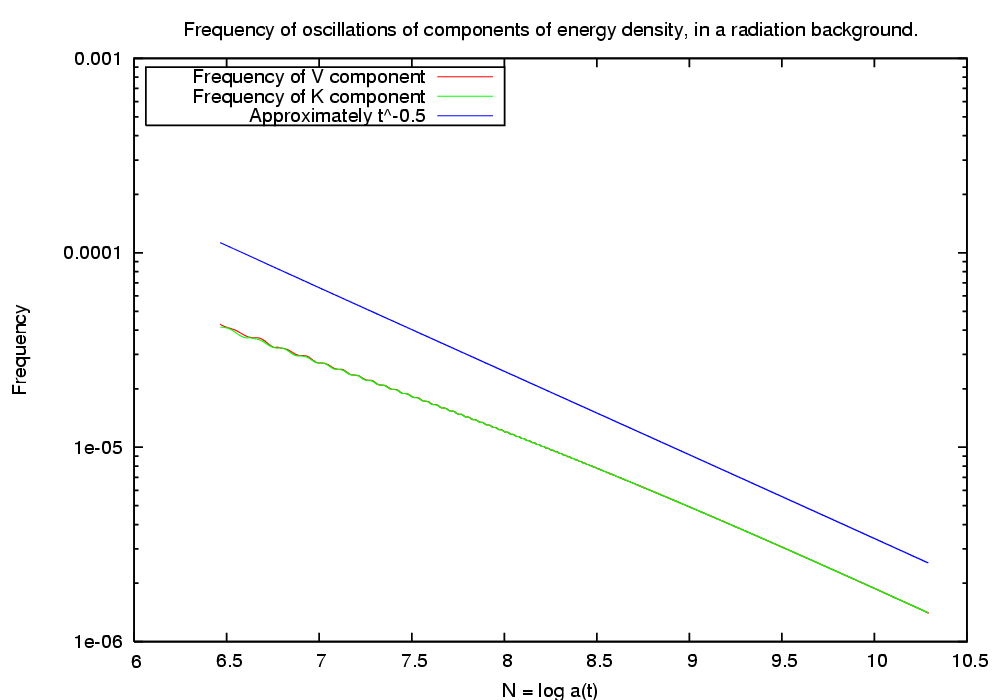}
\flushleft
\caption{The effective frequency of oscillations of the $x$ and $y$ autonomous
system variables \eqref{eqn:aut:vars}.}
\label{fig:canon:scaling:freq}
\end{figure}

The autonomous system analysis we have used is not designed to function in this 
kind of oscillating picture, however if we imagine that it did and that the 
oscillations are caused by a second transverse massive scalar, we might be 
interested in how the frequency associated with that scalar is evolving.  We 
can calculate that from the numerical results utilising an idea from 
\cite{Fodor2006} by which we determine the pointwise wavelength, $T = 2\pi / 
\omega$, of a given simply oscillating function by calculating the value of T 
for which the integral
\ba
\int^{t+t_0/2}_{t-t_0/2}
	\left[ f(t - \frac{T}{2}) - f(t + \frac{T}{2}) \right]^2 dt
\ea
is minimised, over some suitable width $t_0$ around the point in question.
In this way we calculate the frequency of the oscillations in the $x$ and $y$
autonomous variables, picking $t_0$ to be the difference between the pair of
maxima adjacent to each considered point, and minimising for T to arrive at the
frequency plots in Fig. \ref{fig:canon:scaling:freq}.

\section{Evolution near the vacuum: scaling with oscillations}
\label{sec:toyModel}

The plots for $A_a$ (Fig. \ref{fig:canon:fields}) make clear that the later evolution
of the system consists
of shape oscillations around the vacuum $A_1=A_2$, $A_3=A_4$,
$A_5=A_6$, with a monotonic increase in the  volume modulus $\varphi_1$. In a
suitable basis of fields we may write an effective Lagrangian 
\begin{equation}
\label{eqn:EffLag}
\Lag = -\half (\del \varphi_1)^2 - \half (\del\chi_A)^2 - \half \mu_A^2\chi_A^2 e^{-\la \ka \varphi_1},
\end{equation}
where $\la = 3\sqrt{2/7}$ and the index $A$ runs over three dynamical shape
moduli which can be thought of as the ratio of the
repeat lengths in the three toroidally compactified planes. In a FRW background
we therefore arrive at the following field equations for these scalars,
\ba
\label{eqn:VarphiEOM}
\ddot\varphi_1 + 3H\dot\varphi_1  & = & \half \la\ka\mu_A^2\chi_A^2 e^{-\la\ka \varphi_1},\\
\label{eqn:ChiEOM}
\ddot\chi_A + 3H\dot\chi_A  & = & - \mu_A^2\ e^{-\la\ka \varphi_1}\chi_A.
\ea
This effective Lagrangian describes shape moduli oscillations with frequency
$\om_A = \mu_Ae^{-\la\ka\varphi_1/2}$, which contribute an effective potential
to the volume modulus of
$\half \la\mu_A^2\langle \chi_A^2\rangle e^{-\la\ka \varphi_1}$. Here $\langle
\chi_A^2\rangle$ is the time average of the shape moduli oscillations, which
are assumed to have a much shorter period than the timescales set by $H^{-1}$
or $\varphi_1/\dot\varphi_1$.

\newcommand{\ti}{t_\textrm{i}}
Let us assume that the amplitude of the shape moduli oscillations
decay as $t^{-\sigma}$ from some time $\ti$; that the dynamics
behave according to a scaling regime so that each term in the equation
of motion scales with the same power of $t$; and that the energy density 
in the fluid is scaling along with the scalar
so that $H = 2/3\gamma t$. In this case there is an approximate
solution to (\ref{eqn:VarphiEOM}) $\varphi_1 = (\al/\ka)\ln(t/\ti) + \varphi_{1,\textrm{i}}$, with
\begin{eqnarray}
\label{eqn:ApproxSolAlpha}
\al & = & \frac{2\la}{3\ga(2 - \ga)}
\frac{\ka^2 \mu_A^2 \langle\chi_{A\textrm{i}}^2\rangle e^{-\la \ka\varphi_{1,\textrm{i}} }}{3H^2_\textrm{i}}, \\
\label{eqn:ApproxSolSigma1}
\sigma & = & 1 - \half \al\la.
\end{eqnarray}

\noindent
Given that the energy density of the scalar fields is
\begin{equation}
\label{eqn:energy}
\rho_s = \half\dot\varphi_1^2 + \half\dot\chi_A^2 + \half \mu_A^2\chi_A^2 e^{-\la\ka \varphi_1},
\end{equation}
we can take a time average to remove the effects of the oscillations, using
$\langle\dot\chi_A^2\rangle={\omega_A}^2\langle\chi_A^2\rangle$
\begin{eqnarray}
\langle\rho_s\rangle= \frac{\al^2}{2\ka^2 t^2} +  \mu_A^2\langle
\chi_A^2\rangle e^{-\la\ka \varphi_1}.
\end{eqnarray}
Using our approximate solution for $\varphi_1$, as well as its equation
of motion \eqref{eqn:VarphiEOM}, this can be re-expressed as
\begin{equation}
\langle\rho_s\rangle= \frac{\al^2}{2\ka^2 t^2} + \frac{2(2-\ga)}{\la\ga}\frac{\al}{\ka^2 t^2},
\end{equation}
and hence for small $\Omega_\phi$, the kinetic energy of $\varphi_1$ is O($\alpha$)
down from the oscillatory contributions and can indeed be neglected.
We then find that we can express the parameter $\al$ describing the rate of
change of $\varphi_1$ as
\begin{equation}
\label{eqn:AlphaEnergy}
\al  =  \frac{2\la}{3\ga(2 - \ga)}
\Omega_\phi,
\end{equation}
and we see that the time averaged energy scales as $t^{-2}$ which is indeed a
scaling solution and our assumptions are consistent.

Now we turn to the solution for $\chi_A$, which we have assumed takes the form
\begin{equation}
\chi_A \propto \Re \left[ t^{-\sigma}\exp\left( -i\int^t
\om_A(t')dt'\right) \right].
\end{equation}
Substituting into Eq.\ (\ref{eqn:ChiEOM}), and comparing real and imaginary parts, we find
\begin{eqnarray}
\frac{\sigma(1-2/\ga) + \sigma^2}{t^2}- \om_A^2 + \mu_A^2e^{-\la\ka\varphi_1} & = & 0, \\
\frac{2(\sigma-1/\ga)\om_A}{t} - \dot\om_A & = & 0. 
\end{eqnarray}
Hence we recover $\om_A \simeq \mu_Ae^{-\la\ka\varphi_1/2}$ which implies $\dot\om_A
= -\half\la\al\om_A/t$, and thus find that
\begin{equation}
\label{eqn:ApproxSolSigma2}
\sigma = \frac{1}{\ga} - \frac14 \la \al.
\end{equation}
Eqs.\ (\ref{eqn:ApproxSolSigma1}) and (\ref{eqn:ApproxSolSigma2}) are in contradiction unless 
\begin{equation}
\al = 4(\ga-1)/\ga\la,
\end{equation}
and so therefore $\al$ should be approximately $1/\la$ in the radiation era and
vanish in the matter era. Computing $\Om_\phi$ we find 
\begin{equation}
\rho_s \simeq \frac{8(\gamma-1)}{\lambda^2\gamma^2}\frac{1}{\ka^2t^2},  
\end{equation}
and hence
\begin{equation}
\Om_s \simeq \frac{6(\gamma-1)}{\la^2}. 
\end{equation}
In the case of radiation we therefore expect that $\Om_s\simeq 0.7$.
Note that about 1/3 of the energy density in the scalar fields comes from the
kinetic energy of $\varphi_1$, which is non-oscillatory.  Looking at the
radiation era plot in Fig. \ref{fig:canon:scaling:components} we see that there
is a substantial non-oscillatory component and that $\Om_\phi \simeq 0.33$,
which is tantalisingly out by a factor of two from the analysis, but at least in the
right order of magnitude.

The frequency plot is also roughly consistent: the prediction is that
the oscillation frequency should decrease as $t^{-\al\la/2}$, which should be
$t^{-\half}$ in the radiation era, not far from the slope at the end of the run
which was measured to be $t^{-0.5}$.

In the matter era the simulations show that $\Om_\phi \simeq 0.03$, although
the prediction is that it should vanish, which we put this down to a second order
effect of unknown origin. Similarly, the frequency should decay as $t^0$ and we
measured $t^{-0.08}$.



\section{Conclusions}

In this paper we have investigated some cosmological aspects of
compactifications of 11D Einstein gravity on an elliptic twisted
torus, which have the nice property of possessing a positive
semi-definite potential with 4D Minkowski minima, and partial fixing
of the moduli.  We find that slow-roll inflation using the potential
is not possible, as the inflationary $\epsilon$ parameter is always
greater than unity.  We also find that there are novel scaling
solutions in Friedmann cosmologies in which the energy density in
the massive moduli tracks that of the background barotropic fluid,
while the volume modulus increases as approximately $t^{0.5}$ in
the radiation era and a very small power of $t$ in the matter era.

It is an interesting question to ask whether such a scaling solution
could be used to alleviate the cosmological moduli problem.  It
seems phenomologically difficult at first sight, as coupling constants
in the low energy theory typically depend on some power of the
volume modulus, and are quite tightly constrained at and after
nucleosynthesis.

It would also be interesting to explore compactifications with
fluxes \cite{Hull2006}, where a Freund-Rubin flux in the space-time
or wrapping some of the internal dimensions may change our conclusions.

\vspace{1cm}
\noindent
{\large\bf Acknowledgements} JLPK and PMS are supported by S.T.F.C.

\vskip 1cm
\appendix{\noindent\Large \bf Appendices}

\section{Reducing the Ricci scalar}
\label{AppReduceRicci}
We choose a higher dimensional metric to consist of a spacetime part and
an internal Lie group manifold part according to
\ba
ds^2&=&e^{2\psi(x)}ds^2_{(1,d-1)}+g_{ij}(x)e^i\otimes e^j,\\\nonumber
    &=&e^{2\psi(x)}\eta_{\mu\nu}e^\mu\otimes e^\nu+g_{ij}(x)e^i\otimes e^j\\\nonumber
    &=&\hat g_{\hat\mu\hat\nu}e^{\hat\mu}\otimes e^{\hat\nu},
\ea
with the co-ordinates on spacetime being represented by $x$ and those on the
internal space by $y$, $\psi(x)$ represents a freedom to choose the spacetime
co-ordinates.
In the following we shall analyse this space using the frame $e^{\hat\mu}=(e^\mu,e^i)$,
note that this is not an orthonormal frame.
In order to find the connection one-forms, $\omega^{\hat\mu}_{\;\;\hat\nu}$, we
need to solve
\ba
d\hat g_{\hat\mu\hat\nu}-\omega^{\hat\rho}_{\;\;\hat\mu}g_{\hat\rho\hat\nu}
                        -\omega^{\hat\rho}_{\;\;\hat\nu}g_{\hat\mu\hat\rho}&=&0\\\nonumber
de^{\hat\mu}+\omega^{\hat\mu}_{\;\;\hat\nu}\wedge e^{\hat\nu}&=&0,
\ea
and the curvature two-forms follow from
\ba
\hat R^{\hat\mu}_{\;\;\hat\nu}&=&d\omega^{\hat\mu}_{\;\;\hat\nu}+
                            \omega^{\hat\mu}_{\;\;\hat\rho}\wedge \omega^{\hat\rho}_{\;\;\hat\nu}.
\ea
We find that the Ricci tensor is given by
\ba
\label{eqn:scaleFactorEOM}
\hat \Ricci_{\mu\nu}&=& \Ricci_{(d)\mu\nu}
  -(d-2)\nabla_\mu\nabla_\nu\psi
  -\eta_{\mu\nu}\nabla_\rho\nabla^\rho\psi
  -(d-2)\eta_{\mu\nu}\nabla_\rho\psi\nabla^\rho\psi
  +(d-2)\nabla_\mu\psi\nabla_\nu\psi\\\nonumber
  &~&-\frac{1}{4}\nabla_\mu g^{ij}\nabla_\nu g_{ij}
  -\half g^{ij}\nabla_\mu\nabla_\nu g_{ij}
  +\half g^{ij}\left(\nabla_\mu g_{ij}\nabla_\nu\psi+\nabla_\nu g_{ij}\nabla_\mu\psi  \right)
  -\half \eta_{\mu\nu}g^{ij}\nabla_\rho g_{ij}\nabla^\rho \psi\\\nonumber
\label{eqn:mixedRicci}
\hat \Ricci_{\mu j}&=&-\half g^{kl}\nabla_\mu g_{km}f^m_{\;\;lj}\\\nonumber
\label{eqn:shapeModuliEOM}
\hat \Ricci_{ij}&=&\tilde \Ricci_{ij}
  +e^{-2\psi}\left(  \half g^{kl}\nabla_\mu g_{ik}\nabla^\mu g_{jl}
                    -\half\nabla_\mu\nabla^\mu g_{ij}
                    +\frac{1}{4}g_{kl}\nabla_\mu g^{kl}\nabla^\mu g_{ij}
                    -\half (d-2)\nabla_\mu \psi\nabla^\mu g_{ij}\right).
\ea
In deriving this we have used the fact that compact Lie groups are unimodular,
giving $f^I{}_{IJ}=0$ \cite{Scherk1979,Cvetic2003,Cho1975}.
$\tilde{\Ricci}_{ij}$ denotes the curvature of the internal space, treating the
$g_{ij}$ as constant and the covariant derivatives, $\nabla_\mu$ are for
the metric $ds^2_{(1,d-1)}$ with their indices raised by $\eta^{\mu\nu}$.
Given the Ricci curvatures above we can see one of the issues related to
the consistency of truncation, namely that there is nothing to source 
${\Ricci}_{\mu j}$ and so it must vanish by the 11D equations of motion.
For the cases we consider, we find that this term does vanish.

We may now trace the above to find the following Ricci scalar
\ba
\hat {\Ricci}&=&{\Ricci}_{\textrm{int}}
           +e^{-2\psi}[ {\Ricci}_{(d)}-2(d-1)\nabla^2\psi-(d-1)(d-2)\nabla_\mu\psi\nabla^\mu \psi
                      -g^{ij}\nabla^2 g_{ij}\\\nonumber
      &~&             -\frac{3}{4}\nabla_\mu g^{ij}\nabla_\mu g_{ij}
                      -(d-2)g^{ij}\nabla_\mu g^{ij}\nabla^\mu\psi
                      -\frac{1}{4}g^{ij}\nabla_\mu g_{ij}g^{kl}\nabla^\mu g_{kl}].
\ea

\noindent Making use of the gauge freedom we choose
\ba
\label{eqn:gaugeChoice}
e^{(d-2)\psi}\sqrt{g_{ij}} = \kappa^D
\ea
showing that the physical volume of the internal space is given by
\ba
{\cal V}_{phys}&=&+{\cal V}_{\textrm{int}} \, e^{(2-d)\psi}.
\ea
This gauge choice enables us to write
\ba
\hat {\Ricci}_{\mu\nu}&=& \Ricci_{(d)\mu\nu}+\frac{1}{2(d-2)}g^{ij}\nabla_\sigma\nabla^\sigma g_{ij}\eta_{\mu\nu}
                    +\frac{1}{4}\nabla_\mu g^{ij}\nabla_\nu g_{ij}
                    +\frac{1}{2(d-2)}\eta_{\mu\nu}\nabla_\sigma g^{ij}\nabla^\sigma g_{ij}\\\nonumber
           &~&      -\frac{1}{4(d-2)}g^{ij}\nabla_{\mu}g_{ij} g^{kl}\nabla_\nu g_{kl}\\\nonumber
       &=& \Ricci_{(d)\mu\nu}+\frac{1}{4}\nabla_\mu g^{ij}\nabla_\nu g_{ij}
            -\eta_{\mu\nu}\nabla^2\psi-(d-2)\nabla_\mu\psi\nabla_\nu\psi\\\nonumber
\hat \Ricci_{ij}&=&\tilde{\Ricci}_{ij}+\half e^{-2\psi}\left[ g^{kl}\nabla_\mu g_{ik}\nabla^\mu g_{jl}
                                                          -\nabla_\mu\nabla^\mu g_{ij}\right]\\\nonumber
\hat \Ricci&=&e^{-2\psi}[ {\cal R}_{(d)}-2(d-1)\nabla^2\psi-g^{ij}\nabla^2 g_{ij}
                       -\frac{3}{4}\nabla_\mu g^{ij}\nabla^\mu g_{ij}
                       -\frac{1}{4(d-2)}g^{ij}\nabla_\mu g_{ij}g^{kl}\nabla^\mu g_{kl}]\\\nonumber
           &~&+\Ricci_{\textrm{int}}.
\ea

\section{Construction of the elliptic twisted torus}
\label{app:elliptic:example}

These groups \cite{Catal-Ozer2006} are related to the ISO(N) group of
isometries in $N$-dimensional space,
and are described by the structure constants $f^a_{\;\;bc}$, where the indices
$a,b,c=0,1,2...$ and the non-zero components are given by

\ba
f^i_{\;\;0j}&=&M^i_{\;\;j},
\ea
with $i,j=1,2,...$.  The matrix $M$ is skew symmetric and real, and we can
choose a basis in full generality in which it takes the form

\ba
M&=&\left(
\begin{array}{ccccc}
0 & m_1 & 0    & 0 & 0\\
-m1 & 0 & 0    & 0 & 0\\
0   & 0 & 0    & m_2 & 0\\
0   & 0 & -m_2 & 0 & 0\\
0   &   &    0 & 0 & \ddots
\end{array}
\right).
\ea
In terms of left-invariant one-forms we have
\ba
d e^a&=&-\half f^a_{\;\;bc}\; e^b\wedge e^c
\ea
and in terms of generators we have
\ba
\label{structureeqn}
[T_b,T_c]&=&f^a_{\;\;bc}\;T_a.
\ea

To understand the algebra first consider the 3D group,
\ba
[T_0,T_1]&=&-m_1T_2,\quad [T_0,T_2]=m_1 T_1,\quad [T_1,T_2]=0\\
de^0&=&0,\quad de^1=-m_1 e^0 \wedge e^2,\quad de^2=m_1 e^0 \wedge e^1,
\ea
which can be represented with the matrices
\ba
T_0&=&\left(
\begin{array}{ccc}
0 & m_1 & 0    \\
-m1 & 0 & 0    \\
0   & 0 & 0    
\end{array}
\right),\quad
T_1=\left(
\begin{array}{ccc}
0 &   0 & 1    \\
0 & 0 & 0    \\
0 & 0 & 0    
\end{array}
\right),\quad
T_2=\left(
\begin{array}{ccc}
0 &   0 & 0    \\
0 & 0 & 1    \\
0 & 0 & 0    
\end{array}
\right)
\ea
which we recognise as the algebra of translations and rotations in
2D Euclidean space, and give the group element
\ba
g(x,y,\theta)&=&\left(
\begin{array}{ccc}
\cos(m_1\theta)  & \sin(m_1\theta) & x/\kappa    \\
-\sin(m_1\theta) & \cos(m_1\theta) & y/\kappa    \\
0                & 0               & 1    
\end{array}
\right),\quad 0<\theta<2\pi/m_1
\ea
from which we may calculate the left-invariant one-forms
\ba
e_L^0&=&d\theta\\
e_L^1&=& \kappa^{-1} [ \cos(m_1\theta) \, dx - \sin(m_1\theta) \, dy ]\\
e_L^2&=& \kappa^{-1} [ \sin(m_1\theta) \, dx + \cos(m_1\theta) \, dy ]
\ea
and the right-invariant one-forms
\ba
e_R^0&=&d\theta\\
e_R^1&=& \kappa^{-1} [ dx - m y \, d\theta ] \\
e_R^2&=& \kappa^{-1} [ dy + m x \, d\theta ].
\ea
Under left-translations, $g(\theta,x,y)\rightarrow g(\alpha,a,b)g(\theta,x,y)$, we have
\ba
\theta&\rightarrow&\theta+\alpha\\
x&\rightarrow& a+x\cos(m_1\theta)+y\sin(m_1\theta) \\
y&\rightarrow& b-x\sin(m_1\theta)+y\cos(m_1\theta)
\ea
while under right translation, $g(\theta,x,y)\rightarrow g(\theta,x,y)g(\alpha,a,b)$, we have
\ba
\theta&\rightarrow&\theta+\alpha\\
x&\rightarrow& x+a\cos(m_1\theta)+b\sin(m_1\theta)\\
y&\rightarrow& y-a\sin(m_1\theta)+b\cos(m_1\theta)
\ea
which allows us to make identifications on the space, thereby making it
compact.

The adjoint action on the generators is
\ba
h^{-1}(0,a,b)T_ah(0,a,b)&=&D(h)_a^{\;\;b} \, T_b\\
h^{-1}(0,a,b)T_0h(0,a,b)&=&T_0-mb \, T_1+ma \, T_2\\
h^{-1}(0,a,b)T_1h(0,a,b)&=&T_1\\
h^{-1}(0,a,b)T_2h(0,a,b)&=&T_2
\ea
giving
\ba
D(h)_0^{\;\;0}&=&1,\;D(h)_0^{\;\;1}=-mb,\;D(h)_0^{\;\;2}=ma,\\
D(h)_1^{\;\;1}&=&1\\
D(h)_2^{\;\;2}&=&1.
\ea
Under right-translations by $g(0,a,b)$ the left-invariant one-forms change by
\ba
e_L^0&\rightarrow& e^0_L\\
e^1_L&\rightarrow& e^1_L + \kappa^{-1} bm \, e^0_L\\
e^2_L&\rightarrow& e^2_L - \kappa^{-1} am \, e^0_L
\ea
which is the adjoint action of $e^a\rightarrow e^bD(h^{-1})_b^{\;\;a}$.

For our case of interest, 7D, we need the structure constants
\ba
\label{tt7dstructure}
f^1_{\;\;02}=m_1,\qquad
f^3_{\;\;04}=m_2,\qquad
f^5_{\;\;06}=m_3
\ea
with the rest either vanishing or given by anti-symmetry.
Note that the zero index still resides on the 7D space, and is not the
time direction. Also note that this algebra is not semi-simple because
the Killing metric, $G_{cd}=f^a_{\;\;bc}f^b_{\;\;ad}$, is not invertible.
Indeed, the only non-zero component is given by
\ba
G_{00}&=&Tr(M^2).
\ea

In terms of the internal space ansatz, $g_{ab}e^a\otimes e^b$, if the space was
being formed by a coset with a continous subgroup we would expect to find
restrictions on available the degrees of freedom due to invariance and
consistency requirements \cite{Mueller-Hoissen1988}, however there is no such
restriction here when dividing by a discrete subgroup as detailed above.

\section{Scherk-Schwarz ansatz and gauge fixing}
\label{app:ss:gauge:freedom}

We adopt a modified ansatz for the metric, given by the the
Scherk-Schwarz form
\ba
ds^2&=&e^{2\psi(x)}ds^2_{(4)}+g_{ab}\nu^a\otimes\nu^b\\
\nu^a&=&e^a-A^a
\ea
where $A^a$ are gauge fields. Upon substituting this into the Riemann scalar
of the action we find an effective theory described by the scalars $g_{ab}$,
charged under the non-Abelian gauge fields $A^a$.
The gauge transformations are
\ba
\delta A^a&=&d\xi^a-f^a_{\;\;bc}\xi^b A^c.\\
\delta g_{ab}&=&-f^c_{\;\;ad}g_{cb}\xi^d-f^c_{\;\;bd}g_{ca}\xi^d.
\ea
and the covariant derivative of the scalars is
\ba
Dg_{ab}&=&dg_{ab}+g_{ac}f^c_{\;\;bd}A^d+g_{bc}f^c_{\;\;ad}A^d.
\ea
Within Scherk-Schwarz compactification we have the following gauge transformations
\ba
\delta g_{00}&=&-2M^i_{\;\;j}g_{0i}\xi^j\\
\delta g_{0i}&=&-M^j_{\;\;k}g_{ji}\xi^k+M^j_{\;\;i}g_{0j}\xi^0\\
\delta g_{ij}&=&(M^k_{\;\;i}g_{kj}+M^k_{\;\;j}g_{ki})\xi^0
\ea
We may use the $\xi^i$ gauge parameters to set the $g_{0i}=0$ gauge, leaving
us one more gauge degree of freedom which we shall not fix. 
However this does require some assumptions on the rank of $M$.

This all amounts to allowing us to use the gauge $g_{0i}=0$, with no loss of
generality.

\bibliography{twistedpaper}

\end{document}